\documentstyle[sprocl,epsf,epsfig]{article}

\def\beq{\begin{equation}}
\def\eeq{\end{equation}}
\def\bea{\begin{eqnarray}}
\def\eea{\end{eqnarray}}
\def\bq{\begin{quote}}
\def\eq{\end{quote}}

\def\AP{{\it Ann.Phys.} }

\def\NP{{\it Nucl.Phys.} }
\def\PL{{\it Phys.Lett.} }
\def\PR{{\it Phys.Rev.} }
\def\PRL{{\it Phys.Rev.Lett.} }

\def\PTP{{\it Progr.Theor.Phys.} }

\def\gappeq{\mathrel{\rlap {\raise.5ex\hbox{$>$}}
{\lower.5ex\hbox{$\sim$}}}}

\def\lappeq{\mathrel{\rlap{\raise.5ex\hbox{$<$}}
{\lower.5ex\hbox{$\sim$}}}}

\begin{document}
\pagestyle{empty}
\begin{flushright}
CERN-TH/96-284\\
UMN-TH-1517/96\\
TPI-MINN-96/19\\
hep-ph/9610410
\end{flushright}
\vspace*{5mm}
\begin{center}
{\bf ACCELERATOR CONSTRAINTS ON NEUTRALINO DARK MATTER$^*$}\\
\vspace*{0.5cm} 
{\bf J. Ellis}\\
TH Division, CERN, CH-1211 Geneva 23\\
\vspace{0.3cm}
{\bf T. Falk and K.A. Olive}\\
School of Physics and Astronomy, University of Minneapolis,\\
Minneapolis, MN 55455, USA\\
\vspace{0.3cm}
{\bf  M. Schmitt}\\
PPE Division, CERN, CH-1211 Geneva 23\\
\vspace*{0.5cm}  
{\bf ABSTRACT} \\ \end{center}
\vspace*{5mm}
\noindent
The constraints on neutralino dark matter $\chi$ obtained from
accelerator searches at LEP, the Fermilab Tevatron and elsewhere
are reviewed, with particular emphasis on results from LEP 1.5.
These imply within the context of the
minimal supersymmetric extension of the Standard Model that 
$m_{\chi} \ge 21.4$ GeV if universality is assumed, and yield for
large tan$\beta$ a significantly
stronger bound than is obtained indirectly from Tevatron limits
on the gluino mass. We
update this analysis with preliminary results from 
the first LEP 2W run, and 
also preview the prospects for future 
sparticle searches at the LHC.

\vspace*{0.3cm}
\noindent 
\rule[.1in]{12cm}{.002in}

\noindent
${}^*$ Presented by J.E. at the Workshop on the Identification of the Dark
Matter, Sheffield, September 1996.
\vspace*{0.3cm}

\begin{flushleft}
CERN-TH/96-284\\
UMN-TH-1517/96\\
TPI-MINN-96/19\\
October 1996
\end{flushleft}
\vfill\eject

\setcounter{page}{1}
\pagestyle{plain}

\title{ACCELERATOR CONSTRAINTS ON NEUTRALINO DARK MATTER\footnote{
Presented by J.E. at the Workshop on the Identification of the Dark
Matter,\\ Sheffield, September 1996.
}}

\author{J. Ellis} 

\address{TH Division, CERN, CH-1211 Geneva 23, Switzerland}

\author{T. Falk}

\address{School of Physics and Astronomy, University of Minneapolis,
Minneapolis, MN 55455, USA}

\author{K.A. Olive}

\address{School of Physics and Astronomy, University of Minneapolis,
Minneapolis, MN 55455, USA}

\author{M. Schmitt}

\address{PPE Division, CERN, CH-1211 Geneva 23, Switzerland}

\maketitle\abstract{
The constraints on neutralino dark matter $\chi$ obtained from
accelerator searches at LEP, the Fermilab Tevatron and elsewhere
are reviewed, with particular emphasis on results from LEP 1.5.
These imply within the context of the
minimal supersymmetric extension of the Standard Model that 
$m_{\chi} \ge 21.4$ GeV if universality is assumed, and yield for
large tan$\beta$ a significantly
stronger bound than is obtained indirectly from Tevatron limits
on the gluino mass. We
update this analysis with preliminary results from 
the first LEP 2W run, and 
also preview the prospects for future 
sparticle searches at the LHC.}

\section{Theoretical Framework}

We work within the context of the minimal supersymmetric extension
of the Standard Model (MSSM)~\cite{EFOS5}, 
whose gauge interactions are the same
as those in the Standard Model, and whose Yukawa interactions are
derived from a superpotential that conserves $R$ parity and
includes a term that mixes the
two Higgs superfields: $\mu H_1 H_2$. We presume that the lightest
supersymmetric particle (LSP) is a neutralino $\chi$, namely, 
the lightest of the mixtures $\chi_i: i = 1, ...,4$ of the
$U(1)$ gaugino $\tilde B$, the neutral $SU(2)$ gaugino $\tilde W_3$,
and the two neutral Higgsinos $\tilde H_{1,2}$, found by diagonalizing
the mass matrix~\cite{EFOS7}:

\beq
\left(
\matrix{
M_2 & 0 & {-g_2v_2\over\sqrt{2}} & {g_2v_1\over\sqrt{2}} \cr\cr
0 & M_1 & {g^\prime v_2\over\sqrt{2}} & {-g^\prime v_1\over\sqrt{2}} \cr\cr
{-g_2 v_2\over\sqrt{2}} & {g^\prime v_2\over\sqrt{2}} & 0 & \mu \cr\cr
{g_2 v_1\over\sqrt{2}} & {-g^\prime v_1\over\sqrt{2}} & \mu & 0}
\right)
\label{massmatrix}
\eeq
where $g_2, g'$ are the $SU(2)$ and $U(1)$ gauge couplings, 
$v_{1,2} = <0|H_{1,2}|0>$ are the Higgs vacuum expectation values
whose ratio we denote by tan$\beta = v_2/v_1$,
and $M_{1,2}$ are the soft supersymmetry-breaking $U(1)$ and $SU(2)$
gaugino masses. The mass matrix for the charginos $\chi^{\pm}$,
which are mixtures of the charged winos $\tilde W^{\pm}$ and
Higgsinos $\tilde H^{\pm}$ are also characterized by $g_2, v_{1,2}$ 
and $M_2$~\cite{EFOS7}.
We make here the conventional universality assumption that
$M_1 = M_2 \equiv m_{1/2}$ at the supersymmetric GUT 
scale, so that their physical values are renormalized~\cite{EFOS5}:
\begin{equation}
M_2 \, : \, M_1 \, : \, m_{1/2} \, = \, \alpha_2 \, : \, \alpha_1 \, : \,
\alpha_{GUT}
\label{universalratio}
\end{equation}
We also assume universality for the soft supersymmetry-breaking
scalar squared masses: $m^2_{0_i} \equiv m^2_0$ at the 
supersymmetric GUT scale, so that the physical values are
renormalized~\cite{EFOS5}:
\begin{equation}
m^2_{0_i} \simeq m^2_0 + C_i m^2_{1/2} + \hbox{D terms}
\label{scalarratio}
\end{equation}
Theoretically, this assumption is more questionable than
({\ref{universalratio}),
and possible implications of its relaxation are discussed here by
Bottino~\cite{louise}.

\section{Experimental Lower Bound from LEP}

An important experimental step forward in constraining neutralinos was
made possible by the LEP 1.5 run in late 1995~\cite{EFOS1234}. Previously,
searches for
$Z^0 \rightarrow \chi^+ \chi^-$ and $\chi \chi'$ at LEP 1 had not been
able to establish a model-independent lower bound on $m_{\chi}$, as seen
in Fig.~1~\cite{EFOS6}. Nor, indeed, were the LEP 1.5 searches for $e^+
e^- \rightarrow
\chi^+ \chi^-$ and $\chi_i \chi_j$ (whose cross section depends on the
sneutrino mass $m_{\tilde \nu}$) able alone to establish a lower
bound, as also shown in Fig.~1~\cite{EFOS6}. However, the LEP 1.5 
data did serve to fill in a 
`wedge' of parameter space left uncovered by LEP 1 data for $\mu <0$, tan$\beta 
< 2$, as seen in Fig.~2. This was sufficient for the ALEPH
collaboration \cite{EFOS6} to quote a lower limit
\begin{equation}
m_{\chi} \ge 12.8 \, {\rm GeV}
\label{aleph}
\end{equation}
for $m_{\tilde \nu} = 200$ GeV~\footnote{This analysis also excluded
the theoretically-interesting possibility that $\mu = 0$.}. In fact, 
as discussed in \cite{EFOS6} and seen in Fig.~3,
there was still a small loophole for $1 < {\rm tan}\beta < 1.02$ which
could not be excluded by the ALEPH LEP 1.5 data alone, though it could
be excluded by combining them with data from the other LEP
collaborations, or by other considerations \cite{efos}. Of greater
concern was a larger loophole that appeared when $m_0 \sim 60$ GeV and
tan$\beta \sim \sqrt{2}$, as seen in Fig.~4~\cite{EFOS6}, which was due to
a loss of
sensitivity to $\chi^{\pm}$ production because of the invisibility of
$\chi^{\pm} \rightarrow {\tilde \nu} +$ soft $e^{\pm}$ decays made
manifest in Fig.~5~\cite{EFOS6}.

\begin{figure}
%%Figure 1
\hglue2.5cm
\epsfig{figure=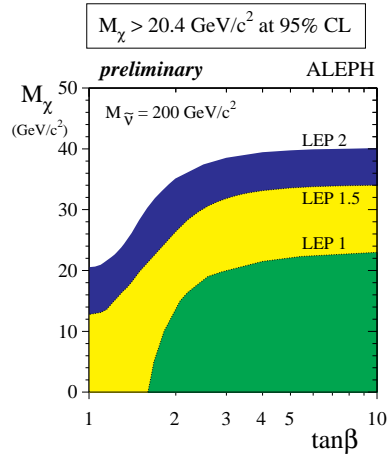,width=5cm}
\caption[]{
Experimental lower bound \cite{EFOS6} on the neutralino mass: note
that neither LEP 1 nor LEP 1.5 data by themselves impose a non-zero lower
bound, though there combination does, modulo the loopholes discussed in the text.
}
\end{figure}

\begin{figure}
%%Figure 2
\hglue2cm
\epsfig{figure=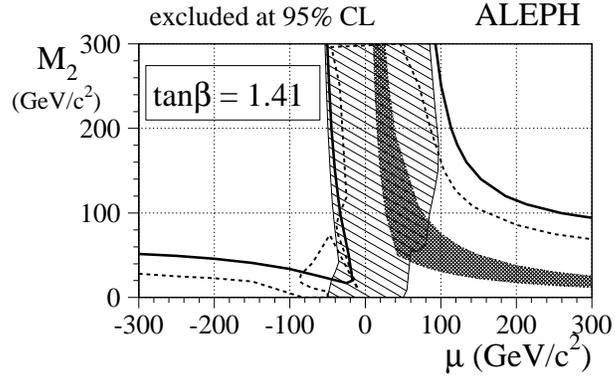,width=8cm}
\caption[]{
The region of the $(\mu , M_2)$ plane excluded by \cite{EFOS6} on the basis 
of searches for charginos and neutralinos at LEP1 and LEP 1.5.
}
\end{figure}

\begin{figure}
%%Figure 3
\hglue2cm
\epsfig{figure=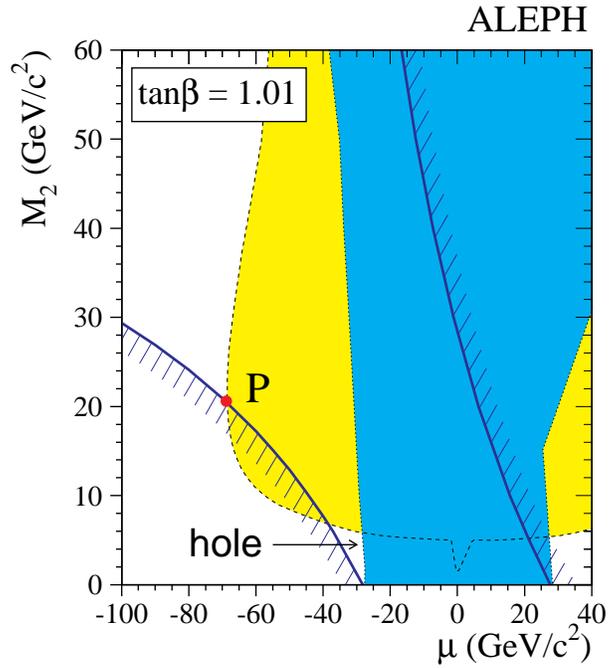,width=8cm}
\caption[]{
The small loophole near $m_{1/2} = 0$ for $1 < \hbox{tan}\beta < 1.02$
in the LEP 1.5 analysis by ALEPH~\cite{EFOS6}.
}
\end{figure}

\begin{figure}
%%Figure 4
\hglue2.5cm
\epsfig{figure=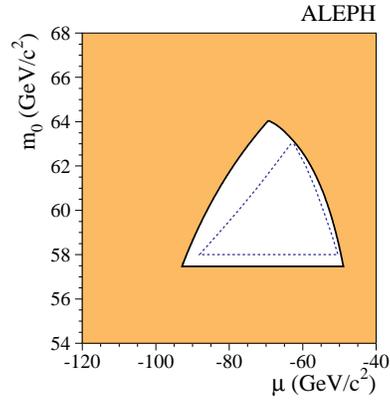,width=5cm}
\caption[]{
The larger loophole for tan$\beta \sim \sqrt{2}$ and $m_0 \sim 60$ GeV
where the ALEPH analysis~\cite{EFOS6} allows $m_{1/2} = 0$.
}
\end{figure}

\begin{figure}
%%Figure 5
\hglue2.5cm
\epsfig{figure=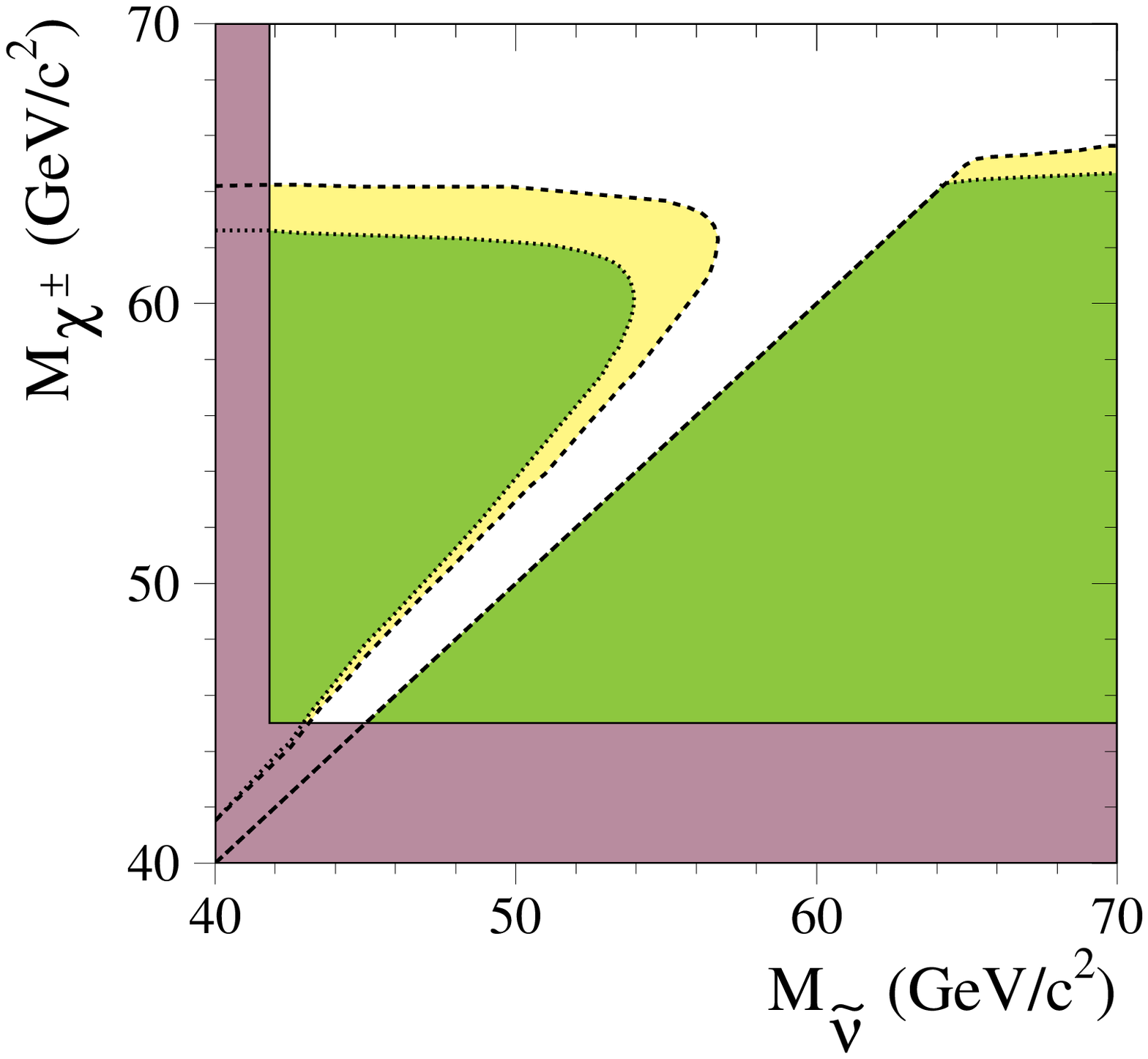,width=5cm}
\caption[]{
The loss of sensitivity in the ALEPH $\chi^{\pm} \rightarrow {\tilde
\nu} + \hbox{soft}~~ e^{\pm}$ search~\cite{EFOS6}, which is responsible for
the loophole shown in Fig.~4.
}
\end{figure}

\section{Phenomenological Analysis}

We~\cite{efos} have attempted to eliminate these two loopholes and
strengthen the
ALEPH lower bound on $m_{\chi}$ by supplementing the Aleph
analysis~\cite{EFOS6} with additional phenomenological inputs. For example,
the neutrino counting analysis at the $Z^0$ peak not only constrains
$Z^0 \rightarrow \chi \chi$ decay, but also $Z^0 \rightarrow {\tilde \nu}
{\bar {\tilde \nu}}$ decay. Taking $N_{\nu} = 2.991 \pm
0.016$~\cite{EFOS14}, we
found that $m_{\tilde \nu} > 43.1$ GeV, if three degenerate flavours of
neutrinos are assumed, as expected in the MSSM with universality. 
Also, LEP 1.5 established new lower limits on charged slepton
masses~\cite{EFOS1234}. As
seen in Fig.~6 for the case tan$\beta = \sqrt{2}$, these
two constraints between them limit the loophole allowing $m_{\chi} = 0$
when $m_{\chi^{\pm}} > m_{\tilde \nu}$, but do not exclude it. However,
this possibility {\it is} excluded by searches at lower centre-of-mass
energies for $e^+ e^- \rightarrow \gamma +$ nothing by the AMY and other
experiments~\cite{EFOS10}, which can be interpreted as upper limits on $\chi
\chi$ production mediated by selectron exchange~\cite{EFOS151617}.
These exclude a zone in the $(m_{1/2}, m_0)$ plane which finally
eliminates
the possibility that $m_{\chi} = 0$, as demonstrated in
Fig.~6~\cite{efos}.

\begin{figure}
%%Figure 6
\hglue2cm
\epsfig{figure=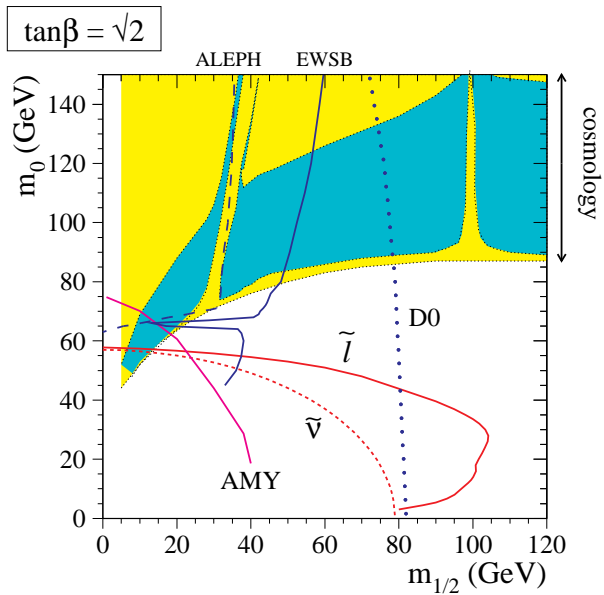,width=8cm}
\caption[]{
The domain of the $(m_{1/2}, m_0)$ plane for $\mu < 0$ and
tan$\beta = \sqrt{2}$ that is excluded by ALEPH chargino and
neutralino searches~\cite{EFOS6} (long-dashed line), by the $Z^0$
limit on $m_{\tilde \nu}$ (short-dashed line), by the LEP
limits~\cite{EFOS1234} on slepton production (solid line), by
single-photon measurements~\cite{EFOS10} (grey line), and by the D0 limit
on the gluino mass~\cite{EFOS24} (dotted line). the region of the plane in
which $0.1 < \Omega_{\chi} h^2 < 0.3$ for some experimentally-allowed
value of $\mu < 0$ is light-shaded, whilst the dark-shaded region is for
$\mu$ determined by dynamical EWSB. the constraint derived from the
ALEPH searches~\cite{EFOS6} when dynamical EWSB is imposed is also shown
as a solid line~\cite{efos}.
}
\end{figure}

\par
To go further, one must take account other phenomenological constraints.
Since we are interested in neutralino dark matter~\cite{EFOS7}, it is
natural to
impose with first priority
the requirement that the relic cosmological density $\rho_{\chi}$
lie in a range of interest to astrophysicists. We base our analysis
on theories of structure
formation based on inflation, with total mass density $\Omega \simeq 1$.
Models with mixed hot and cold dark matter and a flat spectrum of
primordial perturbations, with a cosmological constant and cold
dark matter, and with cold dark matter and a tilted perturbation
spectrum, all favour the range~\cite{EFOS19}
\begin{equation}
0.1 \, \le \, \Omega_{\chi}h^2 \, \le \, 0.3
\label{range}
\end{equation}
which we select for our analysis.

\par
The relic $\chi$ density is controlled by annihilations via ${\tilde
  q}, {\tilde \ell}, Z^0$, neutralino, chargino and Higgs
exchanges~\cite{EFOS18}. Their general trend is to favour some range
of $m_0$ which depends on $\mu$ and $m_{1/2}$ for any given value of
tan$\beta$~\cite{ER}, as illustrated in Fig.~7~\cite{efos}. Note,
however, that this trend is punctuated by holes due to annihilation
via direct-channel $Z^0$ and Higgs poles, which are most important
when $m_{\chi} \sim M_{Z,H}/2$. Over a wide range of $m_{\chi}$, these
cannot be neglected, and require a careful treatment that goes beyond
a simple power-series expansion in the $\chi$ momenta~\cite{EFOS20}.
Fig.~6 displays as the light-shaded region the constraint imposed by
the cosmological density requirement (\ref{range}) in the $(m_{1/2},
m_0)$ plane for tan$\beta = \sqrt{2}$~\cite{efos}.  We see that it
tends to keep $m_0$ away from the dangerous region where $m_{\tilde
  \nu} \lappeq m_{\chi^{\pm}}$, without eliminating it entirely.

\begin{figure}
%%Figure 7
\hglue2.5cm
\epsfig{figure=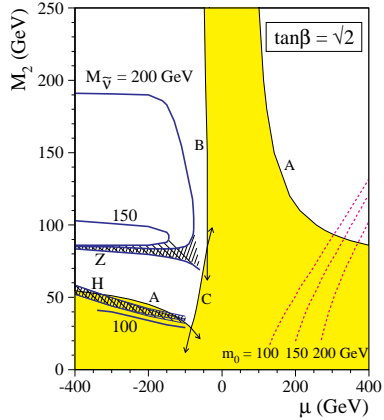,width=5cm}
\caption[]{
The region of the $(\mu, M_2)$ plane excluded by direct
searches~\cite{EFOS1234} for (A) charginos at LEP 1.5, (B)
neutralinos at LEP 1.5 and (C) $Z^0$ decays into $\chi \chi'$
at LEP 1 for tan$\beta = \sqrt{2}$ are indicated by thin solid lines.
Contours of $m_{\tilde \nu}$ (in GeV) required in the MSSM to obtain
$\Omega_{\chi} h^2 = 0.2$ for $\mu < 0$ are indicated by thick solid
lines. The hatched regions indicate where the $Z^0$ and Higgs poles
suppress the relic density. Values of $\mu$ required by dynamical
EWSB for the indicated values of $m_0$ (in GeV) are shown as
short-dashed lines for $\mu < 0$: identical values would be required
for $\mu < 0$~\cite{efos}.
}
\end{figure}

\par
So far, we have not introduced any 
further theoretical assumptions into the MSSM,
beyond those of universality for the scalar and gaugino masses.
It is attractive to hypothesize that electroweak symmetry breaking
(EWSB) is driven dynamically by the renormalization-group running of the
soft supersymmetry-breaking mass of the Higgs boson coupled to
the top quark~\cite{EFOS11}. This EWSB assumption may be regarded
effectively as
fixing $\mu$ for given values of the other MSSM parameters~\cite{EFOS21},
as
illustrated on the right-hand side of Fig.~7 for tan$\beta = \sqrt{2}$,
tending to bound it away from the dangerous regions. In particular,
note that the EWSB assumption cannot be implemented 
for any value of $\mu$ when tan$\beta \lappeq
1.2$ for $m_t \ge 161$ GeV as indicated by experiment, which excludes
the small loophole for tan$\beta \le 1.02$ mentioned earlier~\cite{efos}.
The EWSB assumption may be implemented either in isolation or in
combination with the cosmological constraint (\ref{range}), as seen in
Fig.~6. Taken in isolation, EWSB also reduces the extent of the
loophole where $m_{\tilde \nu} \lappeq m_{\chi^{\pm}}$, without
eliminating it completely. However, cosmology (\ref{range}) in
combination with EWSB is considerably more stringent. The channels
through the darker-shaded region in Fig.~7~\cite{efos} reflect the
positions of the
direct-channel Higgs and $Z$ poles, whose locations are strongly
constrained in this case. Because of the immobility of these channels,
the upper limit in (\ref{range}) on the cosmological density provides
an upper limit on $m_0$ for generic values of $m_{1/2}$, which was not the
case before the imposition of EWSB.

\begin{figure}
%%Figure 8
\hglue2.5cm
\epsfig{figure=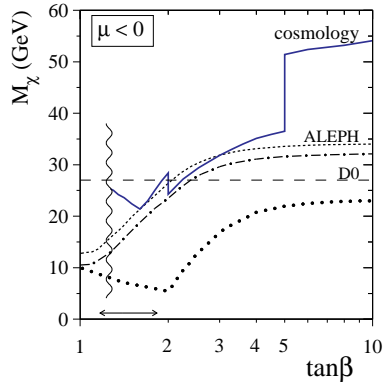,width=5cm}
\caption[]{
The ALEPH lower limit on $m_\chi$ \cite{EFOS6} for $\mu < 0$ and for large 
$m_{\tilde\nu}$ (short-dashed line) is compared, as a function of $\tan\beta$,
with the results obtained in the text by making different phenomenological and
theoretical inputs. The dotted line is obtained by combining the AMY constraint
\cite{EFOS10} with other unsuccessful searches for sleptons and sneutrinos: it excludes
the region of $\tan\beta$, indicated by a double arrow, where the ALEPH 
experimental limit does not exclude $m_\chi$ = 0. The dash-dotted line is
obtained by requiring also that the cosmological relic neutralino density fall
within the preferred range \cite{EFOS19}. The solid line is obtained by combining these
experimental and cosmological inputs with the assumption \cite{EFOS11} of dynamical
electroweak symmetry breaking. The vertical wavy line indicates the lower limit
on $\tan\beta$ in such dynamical electroweak symmetry breaking models. The
horizontal long-dashed line is that obtained from the D0 gluino search \cite{EFOS24},
assuming gaugino mass universality.
%Caption de Fig. 1 de EFOS: ref [n] \rightarrow {EFOSn}, eq. (6)
%\rightarrow eq. (5).
}
\end{figure}

\par 
As an example of the application of the above constraints, let us
consider the specific case tan$\beta = \sqrt{2}$~\cite{efos}, for which
LEP 1 alone
allowed $m_{\chi} = 0$. The ALEPH analysis for $m_{\tilde \nu} = 200$
GeV~\cite{EFOS6}, which is not a conservative assumption, as can be seen
from the
figures, yielded $m_{\chi} \gappeq 17$ GeV. If we relax this
assumption so as to allow any value of
$m_{\tilde \nu}$, but implement all the other experimental constraints
especially that from $e^+ e^- \rightarrow \gamma +$ nothing, we find
$m_{\chi} \gappeq 5$ GeV. This lower bound can be strengthened by
requiring the cosmological constraint (\ref{range}), which yields
$m_{\chi} \gappeq 16$ GeV, modulo a small fraction of the previous
experimental loophole. Finally, if we combine cosmology with the
assumption of dynamical EWSB, we find $m_{\chi} \gappeq 24$
GeV~\cite{efos}.

\par
Our conclusions for general tan$\beta$ are summarized in Fig.~8.
We find that the limit $m_{1/2} \rightarrow 0$ is excluded, as
well as the limit $\mu \rightarrow 0$. We find an absolute lower
limit~\cite{efos}
\begin{equation}
m_{\chi} \, \ge \, 21.4 \, \hbox{GeV}
\label{efoslimit}
\end{equation}
which is attained for tan$\beta \simeq 1.6$. We see in Fig.~8
that, for generic values of tan$\beta$,
this LEP bound is stronger than what could be inferred,
assuming gaugino mass universality, from
the unsuccessful D0 search for gluinos $\tilde g$~\cite{LR}.
This improvement is particularly marked for
large values of tan$\beta$, and is also significant for small
tan$\beta$, particularly if LEP constraints on supersymmetric
Higgs boson masses are taken into account~\cite{EFOS6}, at the price of
additional
sensitivity to theoretical assumptions.

\par
The conclusion (\ref{efoslimit}) has potentially-important
implications for the design of direct experimental searches for
supersymmetric dark matter. It diminishes the priority of
a sensitivity to low $m_{\chi} \lappeq 10$ GeV~\cite{EFOS8}, and it
indicates
that higher nucleon recoil energies may have a higher {\it a priori}
probability. Taken together, these observations indicate that one
might be prepared to sacrifice a lower threshold recoil energy on
the altar of a larger detector mass.

\section{Update Including Preliminary LEP 2W Results}

\par
During the summer of 1996, LEP was run for the first time at
an energy above the $W^+ W^-$ threshold: $E_{cm} = 161$ GeV,
which we term LEP 2W.
The first results of searches during this run
for supersymmetric particles were presented at the Warsaw
ICHEP~\cite{Warsaw} and Minneapolis DPF~\cite{DPF} 
conferences~\footnote{It was commented in the first paper in~\cite{DPF}
that general features of the ALEPH 1.5 analysis~\cite{EFOS6}  were
insensitive to moderate violations of universality between $M_{1,2}$.},
and preliminary
summaries of their analyses have now been presented at CERN
by all the LEP collaborations~\cite{jamboree}. These have included new
upper limits
on chargino, neutralino and slepton production, implying for
example a new lower limit
\begin{equation}
m_{\chi^{\pm}} \, \gappeq \, 80 \times f(\mu, m_{\tilde \nu},
\hbox{tan}\beta)
\, \hbox{GeV}
\label{lep2chargino}
\end{equation}
This and the new preliminary upper limits on $\sigma(e^+ e^-
\rightarrow \chi_i \chi_j)$ can be used to establish a new
preliminary exclusion domain in the $(\mu, m_{1/2})$
plane~\cite{jamboree}, as
shown in Fig.~9. This enables the previous purely experimental
lower limit (\ref{aleph}) to be strengthened to
\begin{equation}
m_{\chi} \gappeq 20 \, \, \hbox{GeV}
\label{lep2neutralino}
\end{equation}
for the case $m_{\tilde \nu} = 200$ GeV, as shown in Fig.~1.

\begin{figure}
%%Figure 9
\hglue2cm
\epsfig{figure=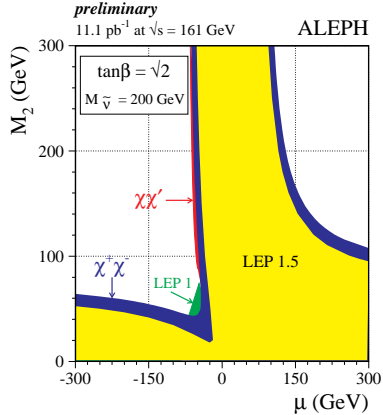,width=5cm}
\caption[]{
The region of the $(\mu, M_2)$ plane excluded by the preliminary
analysis of searches for charginos and neutralinos at LEP
2W~\cite{jamboree}.
}
\end{figure}

Furthermore, this limit does not vary much for $m_{\tilde \nu} \gappeq 80$
GeV. Moreover, the two loopholes where $M_2 = 0$ was formerly
possible, for $1 < \hbox{tan}\beta < 1.02$ at large $m_{\tilde \nu}$
and for tan$\beta \sim \sqrt{2}$ and $m_0 \sim 60$ GeV, are now both
closed by preliminary  LEP 2W data alone~\cite{jamboree},
without the need to
combine them with data from other experiments or to use supplementary
theoretical assumptions.

\begin{figure}
%%Figure 10
\hglue2.5cm
\epsfig{figure=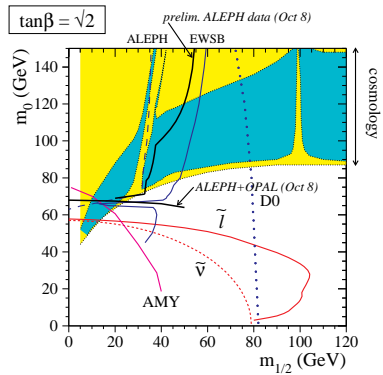,width=5cm}
\caption[]{
Update of Fig.~6, including rough estimates of the potential impact of
the LEP 2W searches for charginos, neutralinos and sleptons. These
estimates represent our personal assessments of preliminary
data~\cite{jamboree}, which will be refined~\cite{efos2} as and when these
data are published.
}
\end{figure}

\par
We have embarked on an improved phenomenological analysis~\cite{efos2},
with the aim of seeing how much the bound (\ref{lep2neutralino}) may be
strengthened by combining the full set of 1996 LEP 2 data with the
cosmological and dynamical EWSB assumptions invoked earlier. Fig.~10
displays a rough assessment of the impact of the preliminary LEP 2W
data on our previous analysis of the excluded domains in the
($m_0, m_{1/2}$) plane shown in Fig.~6. We see that the chargino and
neutralino limits do not by themselves improve significantly the
previous absolute lower limit on $m_{1/2}$, even if our cosmological.
assumption (\ref{range}) is invoked. However, the LEP 2W slepton
limits~\cite{jamboree}
do represent significant new constraints. We have not yet implemented
dynamical EWSB in this updated analysis, in which we plan to include
constraints from searches for supersymmetric Higgs bosons~\cite{jamboree},
which may be significant at low tan$\beta$~\cite{EFOS6}.

\section{Prospects for LHC Searches}

\par
The LHC is designed to have a centre-of-mass energy of $14$ TeV
for $pp$ collisions, at a luminosity ${\cal L} \simeq 10^{34}$
cm$^{-2}$s$^{-1}$, enabling it to explore physics at energy
scales $\lappeq 1$ TeV. In particular, detailed calculations 
of the cross sections for the production of supersymmetric
particles~\cite{sigma} are
available, and it seems that the LHC should be able to detect
the pair production of squarks $\tilde q$ and gluinos $\tilde g$
if their masses are $\lappeq 2$ TeV~\cite{TPs}. Using the proportionality
between $m_{\tilde g}$ and $m_{\chi}$ expected on the basis of
gaugino mass universality, this sensitivity corresponds to a
physics reach up to
\begin{equation}
m_{\chi} \, \simeq \, 300 \, \hbox{GeV}
\label{lhcreach}
\end{equation}
thereby covering most of the range of interest for supersymmetric
dark matter experiments.

\par
The primary sparticle signature studied up to now has been the
classic missing-energy signature of LSP emission~\cite{TPs}, which is
expected
to stand out well above the Standard Model and detector backgrounds,
as seen in Fig.~11. Recent studies indicate that this may be
used to give quite an accurate estimate of the lighter of $m_{\tilde q}$
and $m_{\tilde g}$~\cite{Paige}. The potential importance and interest of
cascade
sparticle decays via intermediate states have been apparent for some
time~\cite{BaerTata}, and their signatures, such as $\ell^{\pm}
\ell^{\pm}, 3 \ell$ and
$Z^0 + E_{T_{miss}}$ final states,
are now being studied in greater detail by the ATLAS and CMS
collaborations~\cite{lhccsusy}.

\begin{figure}
%%Figure 11
\hglue2.5cm
\epsfig{figure=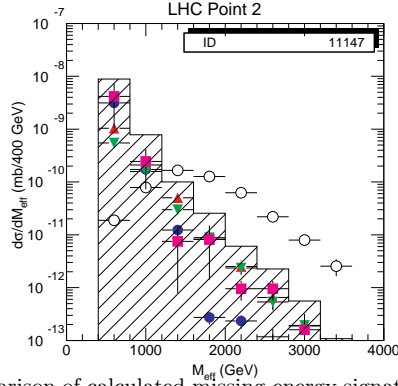,width=5cm}
\caption[]{
Comparison of calculated missing-energy signature due to ${\tilde q},
{\tilde g}$ production at the LHC with Standard Model and
detector backgrounds~\cite{Paige}, in a model with dynamical EWSB and $m_0$ = 
400 GeV, $m_{1/2}$ = 400 GeV, $\tan\beta$ = 10 and $\mu > 0$.
}
\end{figure}

\par
For particular values of the MSSM parameters, cascade decays may
enable the masses of several supersymmetric particles to be
determined simultaneously with high precision~\cite{snowmass}. One generic
possibility is that the cascade includes $\chi_2 \rightarrow \chi +
\ell^+ \ell^-$ decays, which have a sharp end point in $m_{\ell \ell}$,
as seen in Fig.~12~\cite{Paige}. This
may be used to measure $m_{\chi_2} - m_\chi$ with a systematic
uncertainty $\lappeq 50$ MeV! It may then be possible to measure
accurately other sparticle masses by reconstructing the rest of the
decay chain, for example the $\tilde b$ and $\tilde g$ masses in
${\tilde g} \rightarrow {\tilde b} + \chi_2$ decay~\cite{snowmass}. In
this way,
the LHC may be able to measure several combinations of MSSM parameters
with high precision, enabling the relic $\chi$ density to be
calculated more accurately, providing the ultimate accelerator
constraints on neutralino dark matter.

\begin{figure}
%%Figure 12
\hglue3cm
\epsfig{figure=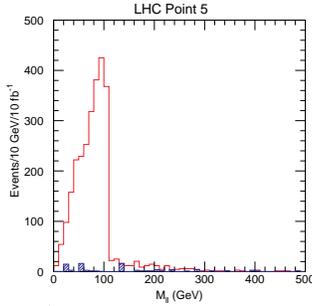,width=4cm}
\caption[]{
Spectrum of $\ell^+ \ell^-$ expected to be produced in $\chi_2
\rightarrow \chi$ decays at the LHC~\cite{Paige}, in a model with
 dynamical EWSB and $m_0$ = 100 GeV,
$m_{1/2}$ = 300 GeV, $\tan\beta$ = 2.1 and $\mu < 0$.
}
\end{figure}

\vfill\eject

\end{document}